\documentclass[preprint,preprintnumbers,amsmath,amssdoes notymb]{revtex4}
\usepackage{amsmath}
\usepackage{amssymb}
\usepackage{bm}
\usepackage{dcolumn}% Align table columns on decimal point
\usepackage{graphicx}
\usepackage{color}%text color

\begin{document}

%%%%%%%%%%%%%%%%%%%%%%%%%
\title{GPGPU for orbital function evaluation with a new updating scheme}
%\title{Quasi-simultaneous updating for QMC electronic structure
%calculations applied to extended solid systems with GPGPU}
%%%%%%%%%%%%%%%%%%%%%%%%%%

\author{Yutaka Uejima}
\author{Ryo Maezono}
\address{School of Information Science, JAIST, Asahidai 1-1, Nomi, Ishikawa, 923-1292, Japan}

\date{\today}

\begin{abstract}
  We have accelerated an {\it ab-initio} QMC electronic structure
  calculation using General Purpose computing on Graphical Processing
  Units (GPGPU).  The part of the code causing the bottleneck for
  extended systems is replaced by CUDA-GPGPU subroutine kernels which
  build up spline basis set expansions of electronic orbital functions
  at each Monte Carlo step.  
  {\color{black} We have achieved a speedup of a factor of
  30} for the bottleneck for a simulation of solid TiO$_2$ with 1,536
  electrons.  To improve the performance with GPGPU we propose a new
  updating scheme for Monte Carlo sampling, quasi-simultaneous
  updating, which is intermediate between
  configuration-by-configuration updating and the widely-used
  particle-by-particle updating.  The error in the energy due to by
  the single precision treatment and the new updating scheme is found
  to be within the required accuracy of $\sim 10^{-3}$ hartree per
  primitive cell.
\end{abstract}

\maketitle

%SSSSSSSSSSSSSSSSSS
\section{Introduction}
%SSSSSSSSSSSSSSSSSS
General Purpose computing on Graphical Processing Units (GPGPU)
\cite{GPU03,GPGPU} has attracted recent interest in HPC (High
Performance Computing) for accelerating calculations at a reasonable
cost.  Environments for developing GPGPU, such as CUDA (Compute
Unified Device Architecture) \cite{CUDA,CUDA2}, have also contributed
to the recent trend for using GPGPU for scientific applications with
much increased portability \cite{GPGPU}.  Electronic structure
calculations \cite{MAR04} form one of the largest fields within HPC
and there have been many attempts to accelerate such calculations
using GPGPU \cite{GPGPU}.  Electronic structure calculation using
quantum Monte Carlo (QMC) methods can provide highly reliable
estimates of material properties for a wide range of compounds
\cite{HAM94,FOU01,NEE10}.  The very high computational demands are not
so important because of the inherently high efficiency of massively
parallel computational facilities for Monte Carlo computations
\cite{GIL11}.  There have been several attempts to apply GPGPU to {\it
  ab-initio} QMC electronic structure calculations
\cite{ESL12,UEJ11}.

\vspace{3mm} Previously we reported GPGPU acceleration of a QMC
calculation for molecular systems, in which we achieved a speedup of
more than a factor of 20 \cite{UEJ11}.  The key idea was to replace
only the bottleneck subroutines in the main code by the CUDA kernel
running on the GPU.  
{\color{black}
We emphasized that the replacement of the entire
simulation code by its GPU version is not practical from the viewpoint
of version administration \cite{UEJ11}.
This becomes more serious for
practical program packages with large number of users, as is common in
{\it ab-initio} electronic structure simulations \cite{UEJ11}.
}

It was challenging to achieve
substantial acceleration using such a `partial replacement strategy',
{\color{black}
and it 
should give a speedup of at least more than a factor of ten 
to be advantageous to use multi-core processor
technology.
}
In Ref.\cite{UEJ11} the main code written in Fortran90 (F90) was partially
replaced by the GPU kernel, which were at the object code level.
Users could switch back to the original CPU version of the subroutine
if the GPU was not available.  In the previous study GPGPU was applied
to molecular simulations, although solid systems are the most
attractive target for GPU-QMC electronic structure simulations
\cite{FOU01} because of the vast CPU time required and the potential
of QMC to achieve more reliable results than frameworks such as
density functional theory (DFT).  

\vspace{3mm} The bottleneck in the present work differs from that in
our previous molecular simulation \cite{UEJ11}.  In our previous work
the bottleneck was the routine for computing the Hartree fields
corresponding to the particle configuration
\cite{MAE07}.  In the present work the bottleneck is the routine for
evaluating the single particle orbitals at the required particle
positions.  We have achieved a speedup of more than a factor of 30
with GPGPU compared with the single core performance of the
conventional CPU evaluation.  This acceleration does not, in
principle, harm the MPI (massively parallel interface) parallelization
efficiency, which is essentially the same as in our previous work
\cite{UEJ11}.  The conventional MPI parallel evaluation \cite{FOU01}
can be accelerated further by attaching a GPU to each node.  In QMC
calculations the electronic orbitals are calculated many times at
different electronic positions.
{\color{black}
It is quite common in {\it ab-initio} electronic structure methods, including
QMC, that one builds up orbital functions for given electronic positions.
Our implementation achieved here would be useful also in self-consistent field
(SCF) methods used in density functional theories (DFT) or molecular orbitals
(MO) methods. 
}

\vspace{3mm} MPI parallelization has successfully been used in QMC
electronic structure calculations \cite{HAM94,FOU01,NEE10}, obtaining
$\sim$ 99\% parallel efficiency by distributing the huge number of
configurations over the processing nodes.  On each node the evaluation
is usually sequential, though there have been several attempts to
exploit further parallelization within the node using, for example,
OpenMP \cite{NEE10}.  The evaluations performed on each node include
updating an electronic configuration $\vec R^{(\alpha)} = (\vec
r_1^{(\alpha)},\cdots,\vec r_j^{(\alpha)}, \cdots,\vec
r_N^{(\alpha)})$, and sampling with the updated configuration, where
$\alpha$ is the index for MC steps.  There are two major types of
updating scheme, the configuration-by-configuration scheme
(simultaneous updating) and the particle-by-particle scheme (PbP,
sequential updating).  In the former, attempted trial $N$-electron
moves are generated to update a configuration, $\vec R^{(\alpha)} \to
\vec R^{(\alpha + 1)}$, and then the new configuration is accepted or
rejected.  In the latter, a trial move of a single electron is
attempted and accepted or rejected, $\vec r_j^{(\alpha)} \to \vec
r_j^{(\alpha +1)}$, and the process is repeated $N$ times.  Sequential
updating is more efficient than simultaneous updating, and it is
widely used in QMC electronic structure calculations \cite{NEE10}.
For hybrid parallelization, including GPGPU and OpenMP, one seeks
further parallelization in the sequential evaluation within an MPI
node.
The GPGPU performs the accept/reject steps for each particle
`simultaneously'.  The ratio of the probabilities evaluated in the
Metropolis accept/reject algorithm \cite{HAM94} differs both from
those for simultaneous and sequential updating.  In this sense our
updating scheme can be viewed as `quasi-simultaneous updating' (Q.S.).
This scheme is designed to obtain GPGPU acceleration by improving the
sequential memory access (so called `coalescing'), and the concealment
of memory latencies.  We have confirmed that our new updating scheme
does not change the results within the required statistical accuracy,
namely the chemical accuracy.

\vspace{3mm} The paper is organized as follows.  In \S II we briefly
summarize the VMC method (Variational Monte Carlo method).  The
evaluation of the orbitals represented in a spline basis set is the
bottleneck in the computation, as described in this section.  The
benchmark systems used in the performance evaluation are also
introduced.  \S III is devoted to {\color{black} a description} of the GPU
architecture.  The structure of processors and memories in the GPU
used in the present work is briefly explained.  Other features, such
as how we assign the number of threads and blocks for parallel
processing, are discussed in \S IV, in connection with the design and
implementation of the quasi-updating scheme.  The quasi-updating
scheme is also introduced in this section.  Several other possible
implementations with different updating schemes or thread/block
assignments are introduced here and their performances are compared.
The results are summarized in \S V, including comparisons of the
energies, operation costs and data transfers, and the dependence of
the performance on system size.  In \S VI we discuss the results,
comparing with the ideal performance in terms of operations and memory
access.  We also discuss the possibility of using high-speed memory
devices in the GPU and the relation to linear algebra packages.

%%%%%%%%%%%%%%%%
\section{QMC electronic structure calculation}
\subsection{VMC}
\label{VMC}
%%%%%%%%%%%%%%%%
In {\it ab-initio} calculations the system is specified by a hermitian
operator $\hat H$ called the Hamiltonian \cite{PAR94}.  The operator
includes information about the positions and charges of the ions, the
number of electrons, and the form of the potential functions in the
system.  The fundamental equation at the electronic level is the
many-body Schr\"odinger equation, which takes the form of a partial
differential equation with $\hat H$ acting on a multivariate function
$\Psi\left(\vec r_{1},\cdots,\vec r_{N}\right)$, known as the
many-body wave function, where $N$ denotes the number of electrons.
The energy of the system, $E$, is the eigenvalue of the partial
differential equation.  The equation has the variational functional
\cite{HAM94}
%eq%eq%eq%eq%eq%eq%eq
\begin{eqnarray}
\label{eq:vmc_energy}
E &=& \frac{\int \Psi^* \hat{H} \Psi \, d{\vec r_1}\cdots d{\vec r_N}}
{\int \Psi^* \Psi \,
d{\vec r_1}\cdots d{\vec r_N}} 
\nonumber \\
&=& 
\frac{\int |\Psi|^2 \cdot\Psi^{-1} \hat{H} \Psi \, d{\vec r_1}\cdots d{\vec r_N}}
{\int |\Psi|^2 \, d{\vec r_1}\cdots d{\vec r_N}} ,
\end{eqnarray}
%eq%eq%eq%eq%eq%eq%eq
which is minimized when the above integral is evaluated with $\Psi$ being an
exact solution of the eigen equation.  For a trial $\Psi$ the functional can
be evaluated as an average of the local energy, $E_L \left(\vec
r_1,\cdots,\vec r_N\right) = \Psi^{-1} \hat{H} \Psi$ over the probability
density distribution
\begin{eqnarray}
  p(\vec r_1,\cdots,\vec r_N) = |\Psi|^2/\int |\Psi|^2 \, d{\vec r_1}
  \cdots d{\vec r_N} \ .
\label{eq05mar12_1}
\end{eqnarray}
In VMC the energy is evaluated by Monte Carlo integration using the
Metropolis algorithm to generate configurations $\left\{\vec
  R^{(\alpha)}\right\}_{\alpha=1}^{M}$ distributed according to the
probability distribution $p(\vec r_1,\cdots,\vec r_N)=p(\vec R)$,
where $\vec R$ denotes a configuration $(\vec r_1,\cdots,\vec r_N)$,
as
%eq%eq%eq%eq%eq%eq%eq
\begin{equation}
  E = \frac{1}{M}\sum_{\alpha=1}^{M} {E_L\left(\vec R^{(\alpha)}\right)} \ ,
\label{eq2}
\end{equation}
%eq%eq%eq%eq%eq%eq%eq
with $M$ being typically of the order of millions.  The trial function
$\Psi$ is improved by an optimization procedure so that the integral
of Eq.\ (\ref{eq:vmc_energy}) can be minimized numerically
\cite{Optimization1,Optimization2}.  Since each $E_L \left( \vec
  R^{(\alpha)} \right)$ is evaluated independently, the summation over
$\alpha$ can be divided into sub-summations distributed over the
processors by MPI with high efficiency \cite{FOU01}.  In this work
GPGPU is used to accelerate the evaluation of each $E_L \left( \vec
  R^{(\alpha)} \right)$, rather than parallelization over the suffix
$\alpha$.  We used the `CASINO' program package \cite{NEE10} for the
VMC calculations.

\vspace{3mm} There are several possible forms of trial $\Psi$, and we chose to
use the common Slater-Jastrow type wave function \cite{FOU01,HAM94},
%eq%eq%eq%eq%eq%eq%eq
\begin{equation}
\label{eq6}
\Psi _{\rm SJ} \left( {\vec R} \right) = e^{J\left( {\vec R} \right)}  
\cdot \Psi _{\rm D} \left( {\vec R} \right)
\ ,
\end{equation}
%eq%eq%eq%eq%eq%eq%eq
\noindent
where $e^{J\left( {\vec R} \right)}$ is known as the Jastrow factor
\cite{JAS55,Neil_jastrow}.  $\Psi_{\rm D}$ is a Slater determinant
\cite{ASH76}
\begin{equation}
\label{eq5}
\Psi_{\rm D} \left( {\vec r_1 , \cdots ,\vec r_N } \right) 
= \left| {\begin{array}{*{20}c}
   {\psi _1 \left( {\vec r_1 } \right)} &  \cdots  
   & {\psi _N \left( {\vec r_1 } \right)}  \\
    \vdots  &  \ddots  &  \vdots   \\
   {\psi _1 \left( {\vec r_N } \right)} &  \cdots  
   & {\psi _N \left( {\vec r_N } \right)}  \\
\end{array}} \right|
\ ,
\end{equation}
\noindent
which is an anti-symmetrized product of one-particle orbital
functions, $\left\{\psi _l \left( {\vec r} \right)\right\}_{l=1}^L$.
The number of independent orbitals, $L$, can be reduced by using the
symmetries of the system.  The bottleneck of the whole simulation has
been found to be the construction of the $\left\{\psi _l \left( {\vec
      r} \right)\right\}$ \cite{NEE10}.  In this study the
computational power of the GPU is devoted to the bottleneck process,
as described in the following subsection.

%SSSSSSSSSSS
\subsection{Orbital evaluation}
\label{Orbital evaluation}
\begin{figure}[h]
\begin{center}
\includegraphics[width=90mm,angle=0]{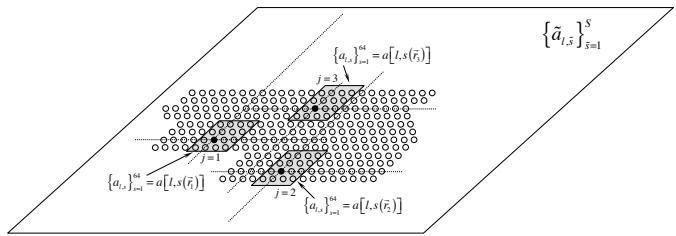}
\end{center}
\caption{ Data structure for the expansion coefficients of the orbital
  functions in Eq.\ (\ref{11.12.24.1}), schematically depicted in two
  dimensional space.  Black lattice points show the nearest site for
  each given $\vec r_j$.  In the shaded regions the basis functions
  $\left\{\Theta_s \left( {\vec r_j} \right)\right\}$ in Eq.\
  (\ref{11.12.24.1}) have non-zero values. }
\label{coefficient}
\end{figure}

%SSSSSSSSSSS
In each MPI process the following evaluations are performed sequentially:
\begin{enumerate}
\item An attempted move, $\vec R^{(\alpha)} \to \vec R^{(\alpha +
    1)}$, is randomly generated,
\item The updated probability $p(\vec R^{(\alpha + 1)})$ and the ratio
  $\xi = p(\vec R^{(\alpha +1)})/p(\vec R^{(\alpha)})$ is evaluated,
\item Based on the ratio $\xi$, the attempted move $\vec R^{(\alpha + 1)}$ is
  accepted or rejected,
\item The local energy $E_L \left( \vec R^{(\alpha + 1)}\right)$ is evaluated.
\end{enumerate}
Each configuration is a set of electronic positions (we omit the spin
coordinate for simplicity), $\vec R^{(\alpha)} = (\vec
r^{(\alpha)}_1,\vec r^{(\alpha)}_2,\cdots,\vec r^{(\alpha)}_j,
\cdots,\vec r^{(\alpha)}_N)$.  Following Eqs.\ (\ref{eq05mar12_1}),
(\ref{eq6}), and (\ref{eq5}), one can reduce the evaluation of the
ratio $\xi = p(\vec R^{(\alpha +1)})/p(\vec R^{(\alpha)})$ to that of
the orbital functions, $\left\{\psi _l \left( {\vec r^{(\alpha +1)}_j}
  \right)\right\}$.  In practical QMC calculations for extended
systems the orbital functions are expanded in a $B$-spline basis set
$\left\{\Theta_s \left( {\vec r} \right)\right\}$
\cite{ALF04,MAE09,NEE10} as
%eq%eq%eq%eq%eq%eq%eq%eq%eq%eq%eq%eq
\begin{equation}
\label{11.12.24.1}
\psi_l \left( {\vec r_j} \right) 
= \sum\limits_{s = 1}^{4^3=64} {a_{ls} 
\cdot \Theta _s \left( {\vec r_j} \right)}   \ .
\end{equation}
%eq%eq%eq%eq%eq%eq%eq%eq%eq%eq%eq%eq%eq
The index $s$ runs over the subset of the spatial sites within the
unit cell of the periodic system.  The spline basis functions,
$\left\{\Theta_s \left( {\vec r} \right)\right\}$, have non-zero
values only at sites $s$ within the fourth nearest neighbor of the
position $\vec r$ along each direction.  The total number of terms in
the summation (\ref{11.12.24.1}) is therefore 64 = 4$^3$ (four spatial
points along each direction in the three dimensional space), as a
subset of the whole lattice within the unit cell amounting to $S=50^3
\sim 60^3$.  The lattice is indexed by $\left\{\tilde
  s\right\}_{\tilde s = 1}^{S}$, as depicted schematically in the two
dimensional plane in Fig.\ \ref{coefficient}.  The subset
$\left\{s\right\}_{s=1}^{4^3} \subset \left\{\tilde s\right\}_{\tilde
  s = 1}^{S=50^3\sim 60^3}$ is the spatial region where
$\left\{\Theta_s \left( {\vec r} \right)\right\}$ has non-zero values,
depending on the given $\vec r$.  Since the indices introduced so far
are complicated we summarize them in Table \ref{indices}.
%table%table%table%table%table%table%table
\begin{table}[htdp]
\caption{Conventions for indices used in this paper.}
\begin{center}
\begin{tabular}{c||c|c|c}
\noalign{\hrule height 1.5pt}
 & Index & Total Amount & Reference  \\ 
\noalign{\hrule height 1.5pt}
Configurations   & $\alpha$ & $M\sim$ millions & Eq.\ (\ref{eq2})   \\ \hline
Orbitals                & $l$ & $L < N$ & Eq.\ (\ref{eq5}) \\ \hline
Electrons             & $j$ & $N$ & \S II.B  \\ \hline
Blip Grid (subset)         & $s$ & $4^3 = 64$ &   Eq.\ (\ref{11.12.24.1})\\ \hline
Blip Grid (whole)       & $\tilde s$ & $S=50^3\sim 60^3$ &   Eq.\ (\ref{7})\\ 
\noalign{\hrule height 1.5pt}
\end{tabular}
\end{center}
\label{indices}
\end{table}
%table%table%table%table%table%table%table

The value of $\Theta_s \left( {\vec r} \right)$ at a $s$-lattice site, $\vec
R_s = (X_s,Y_s,Z_s)$, is given by the function depending on the distance
between $\vec r$ and $\vec R_s$ as,
%eq%eq%eq%eq%eq%eq%eq%eq%eq%eq%eq%eq%eq%eq%eq%eq%eq%eq%eq%eq%eq%eq%eq%eq%eq
\begin{equation}
\label{11.12.24.2}
\Theta _s \left( {\vec r} \right) 
= \varphi \left( {\frac{{x - X_s}}{b_x}} \right) 
\cdot \varphi \left( {\frac{{y - Y_s}}{b_y}} \right) 
\cdot \varphi \left( {\frac{{z - Z_s}}{b_z}} \right) \  ,
\end{equation}
%eq%eq%eq%eq%eq%eq%eq%eq%eq%eq%eq%eq%eq%eq%eq%eq%eq%eq%eq%eq%eq%eq%eq%eq%eq
where $\varphi \left( \zeta \right)$ is a second order polynomial in $\zeta$,
and $(b_x,b_y,b_z)$ denote grid spacings for each direction.  The coefficients
in Eq.\ (\ref{11.12.24.1}), $\left\{ {{{\tilde a}_{l,\tilde s}}}
\right\}_{\tilde s = 1}^{{S\sim 250,000}}$, are precomputed and provided as an
input file, stored in memory at the beginning of the simulation.  For each MC
step with a updated particle position $\vec r_j$, the subset
%eq%eq%eq%eq%eq%eq%eq
\begin{equation}
\left\{ {\left\{ {{a_{l,s\left( {{{\vec r}_j}} \right)}}} 
\right\}_{s = 1}^{64}} \right\}_{j = 1}^N 
\subset 
\left\{ {{{\tilde a}_{l,\tilde s}}} \right\}_{\tilde s = 1}^{{S}}
\label{7}
\end{equation}
%eq%eq%eq%eq%eq%eq%eq
is identified and used in the summation (\ref{11.12.24.1}).  Denoting
$a_{l,s\left( {\vec r_j } \right)} = a\left[ {l,j_x ,j_y ,j_z }
\right]$ as an array, $s\left( {\vec r_j } \right)=(j_x ,j_y ,j_z)$
forms a simply connected region in the three dimensional space but it
does not allow sequential memory access in one dimensional address
space, as it is interrupted with some stride due to the higher
dimensions (see Fig.\ \ref{coefficient}).  The orbital index $l$, is,
however, inherently one dimensional and we exploit this for sequential
memory access, which is very important for GPGPU, as discussed in \S
IV.B.
%fig%fig%fig%fig%fig%fig%fig%fig
%fig%fig%fig%fig%fig%fig%fig%fig

\vspace{3mm} We have identified the `multiply and add' operations in
Eq.\ (\ref{11.12.24.1}), 
{\color{black}
by which the orbital functions, $\left\{\psi_l \left( {\vec r_j} \right)\right\}_{l=1}^{L}$, are evaluated
}
as the bottleneck in the present QMC simulation \cite{NEE10}.  This operation
appears at every MC step when the particle position is updated, $\vec
r_j^{(\alpha)} \rightarrow \vec r_j^{(\alpha +1)}$.  The number of
operations for a single evaluation is proportional to the number of
orbitals, $L$, and hence to $N$.  In the present study we treat system
sizes up to $L=384$ and $N$=1,536.

%section%section%section%section%section%section%section%section%section%section%section%section%section
\subsection{Benchmark systems}
%section%section%section%section%section%section%section%section%section%section%section%section%section
To investigate the dependence of the acceleration on system size, we
prepared three different benchmark systems, as reported in Table
\ref{system_size}.  For each system the atomic cores are replaced by
pseudopotentials \cite{MAR04} in Si-diamond ($N$=216) and cubic
TiO$_2$ ($N$=648 and 1,536).  The periodic boundary conditions for
$(3\times 3\times 3)$ or $(4\times 4\times 4)$ arrays of unit cells
form a simulation cell.  More detailed specifications for each system
are 
{\color{black}
given in Ref.\
\cite{MAE10} for Si and Ref.\ \cite{MOH12} for TiO$_2$.
}
%table%table%table%table%table%table%table
\begin{table}[htdp]
  \caption{Benchmark systems used in the present study.  $N$ and $L$ denote
    the number of electrons and orbitals for each system, respectively.  The
    timing data and the acceleration factors achieved by the best coding are
    summarized, see \S V.A. }
\begin{center}
\begin{tabular}{c|c|c||c|c|c}
\noalign{\hrule height 1.5pt}
System & $N$ & $L$ & CPU time (ms)& GPU time (ms)& Acceleration factor \\ 
\noalign{\hrule height 1.5pt}
Si ($3\times 3\times 3$)&216  & 56    & 2.77 & 0.17   & 16.58 \\ \hline
TiO$_2$ ($3\times 3\times 3$) & 648   & 168  & 20.21  & 0.82 & 24.46 \\ \hline
TiO$_2$ ($4\times 4\times 4$) & 1,536 & 384  & 100.00 & 3.26 & 30.67 \\ 
\noalign{\hrule height 1.5pt}
\end{tabular}
\end{center}
\label{system_size}
\end{table}
%table%table%table%table%table%table%table

\vspace{3mm} The bottleneck routine (orbital evaluation) to be
replaced by the GPU processing kernel occupies 20$\sim$30\% of the
entire CPU time for TiO$_2$ ($N$=1,536), as analyzed by a profiler
(Intel VTune Amplifier \cite{vtune}).  This depends on the choice of
the `dcorr' parameter in CASINO \cite{NEE10,HON10}, which specifies
the interval between sampling; in order to reduce the correlation in
the sampling, the local energy is evaluated every `dcorr' MC steps (an
MC step corresponds to the update of a configuration).  The ratio of
the CPU time spent in the bottleneck is reduced from 39.5\% to 22.0\%
by increasing `dcorr' from one to ten, as measured for a simulation
with 10,000 MC steps.  Typically `dcorr = 5' is chosen, for which the
reduction becomes 27.5\%.  The reason for this dependence is that the
orbital evaluation is called not only by the configuration updating
but also by the local energy evaluation.  Increasing `dcorr' means
less frequent evaluation of the local energy and hence less frequent
calls to the orbital evaluation.

%section%section%section%section%section%section%section%section%section%section%section%section%section
\section{GPU}
%section%section%section%section%section%section%section%section%section%section%section%section%section
General descriptions of the architecture of a GPU can be found in the
literature \cite{GPGPU,CUDA,CUDA2}, and our previous paper \cite{UEJ11} also
provides such a description.  This section provides the minimum amount of
information required to understand the present work which was performed with
the NVIDIA GeForce GTX 480 architecture.

\vspace{3mm} A GPU has hundreds of processing cores.  The key points
for the acceleration in the present work can be summarized as follows:
(1) Parallelized tasks are distributed over many cores. The large
number of processing cores of a GPU allows the whole task to be
processed more rapidly than by a CPU.  (2) The parallelized tasks are
grouped into several sets (called `warps').  The GPU processes each
warp in order (`barrel processing'), skipping those still waiting for
data load from memory.  As there are usually hundreds of warps, barrel
processing conceals the memory latency.  In our previous work
\cite{UEJ11} the acceleration was achieved mainly by dividing a huge
number of loops into several subsets and distributing them over GPU
processor cores.  The present work does not follow this strategy, and
we do not divide the loop for the summation in Eq.\
(\ref{11.12.24.1}).  
{\color{black}
Instead a huge number of independent parallel
tasks, $N\times L =1,536 \times 384 = 589,824$, for the orbitals
$\left\{\left\{\psi_l \left( {\vec r'_j}
    \right)\right\}_{l=1}^{L}\right\}_{j=1}^N$ are distributed over
the GPU processing cores.
}  Other key points for the present work
include, (3) memory latency is much improved when the access occurs
with sequential memory address (memory coalescing), and (4) 
{\color{black}
a command set called a `Fused Multiply Add (FMA) which 
performs multiply and add operations within a clock cycle (two
operations at once).
}

%%%%%%%%%%%%%%%%%%
\subsection{Processors and performance}
\begin{figure}[h]
\begin{center}
\includegraphics[width=80mm,angle=0]{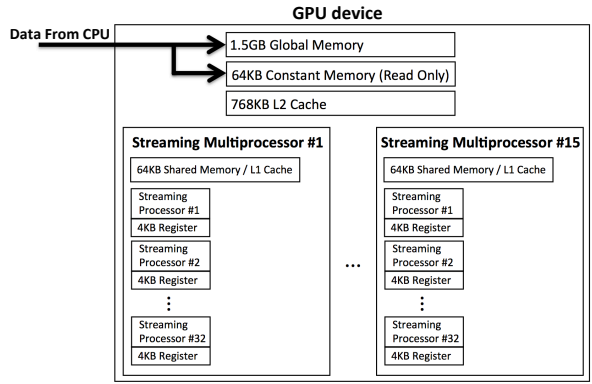}
\end{center}
\caption{A schematic picture of the structure of the GPU device used in the
  present work.  }
\label{MemoryStructure}
\end{figure}

%%%%%%%%%%%%%%%%
GTX480 has 480 processor cores, each of which is termed a `streaming
core' (SP) for AMD products while `cuda core' is used for NVIDIA
products.  In the present paper we use the term SP.  The specs of the
GTX480 are summarized in Table \ref{gpu_spec}.  As shown in Fig.\
\ref{MemoryStructure} every 32 SPs are grouped into a unit called a
Streaming Multi-Processor (SM), the total number of which is hence
fifteen.  Each SP includes 32 bit scalar operators for floating point
(FP32) and integer (Int32) data.  These two operators can process the
data independently within a clock cycle, giving a contribution to the
ideal performance with 2 OP/cycle (two operations per clock cycle) for
single precision operations.  For double precision each 32 bit
operator deals with 64 bit floating point (integer) data with two
clock cycles, termed FP64 (Int64), giving a 1 OP/cycle contribution on
average for double precision operations.  There is another kind of
operation unit called a `Special Function Unit' (SFU), devoted to
evaluating hyper functions including exponential, logarithmic, and
trigonometric functions.  Each SM includes four SFUs in addition to
the 32 SPs.  A SFU performs four floating point operations per clock
cycle, in parallel with other SM operations, which therefore
contributes a further 4 OP/cycle to the ideal performance.
%table%table%table%table%table%table%table%table%table
\begin{table}[htdp]
\caption{Spec of the NVIDIA GTX480 GPU architecture used in the present work.}
\begin{center}
\begin{tabular}{c|c}
\noalign{\hrule height 1.5pt}
Compute Capability &2.0 \\ \hline
Clock of CUDA cores &1401 MHz \\ \hline
Global Memory & 1536 MB \\ \hline
Memory Bandwidth & 177.4 GB/s \\ \hline
Number of SM & 15 \\ \hline
Number of CUDA Cores & 480 \\ \hline
Constant Memory &  64 KB \\ \hline
Shared / L1 Memory & 16 or 48 KB per block\\ \hline
Max number of Threads & 1024 per block \\ 
\noalign{\hrule height 1.5pt}
\end{tabular}
\end{center}
\label{gpu_spec}
\end{table}
%table%table%table%table%table%table%table%table%table
With a clock frequency of 1.401 GHz, the peak performance of GTX480 is
hence evaluated as
\[
{\rm{1.401 GHz}} \times 15{\rm{SM}} \times ({\rm{32SP}} \times {\rm{2OP}} 
+ {\rm{4SFU}} \times {\rm{4OP) = 1,681[GFlops]}} \ .
\]

\vspace{3mm} Note that, unlike the previous GTX275, multiply-and-add
operations are not subject to a SFU in the present GTX480.  For
evaluating Eq.\ (\ref{11.12.24.1}) there is hence no place for a SFU
to be applied, giving an ideal performance to be compared with our
achievement of
\[
{\rm{1.401 GHz}} \times 15{\rm{SM}} \times ({\rm{32SP}} \times {\rm{2OP) 
= 1,345 [GFlops] \ ,
}}
\]
by omitting the contribution from SFU.  Though the present work
concentrates on single precision GPU evaluation, the ideal double
precision performance is estimated to be 672 [GFlops], which is half
of that for single precision.  GeForce GTX480, however, limits it to a
quarter of this value, 168 [GFlops], by its driver, for some reason
\cite{UEJ11}.  These estimates are summarized in Table \ref{cpu_gpu},
compared with that of the CPU (Intel Core i7 920) used in the present
work.

%fig%fig%fig%fig%fig%fig%fig%fig%fig%fig
%fig%fig%fig%fig%fig%fig%fig%fig%fig%fig

%table%table%table%table%table
\begin{table}[htdp]
  \caption{Estimated ideal performances of GTX480 to be compared with our
    achievement.  The peak performance of the Intel Core i7 CPU is also listed
    for reference.  Note that the `ideal' performance differs from the peak
    performance of the device (see the text).  }
\begin{center}
\begin{tabular}{c|c|c|c}
\noalign{\hrule height 1.5pt}
& Clock freq.\ & No. of & Ideal Performance\\ 
& (GHz) & cores & (GFlops) \\ 
\noalign{\hrule height 1.5pt}
GPU (GTX480) Single precision& 1.401 & 480 & 1,345\\ \hline
GPU (GTX480) Double precision& 1.401 & 240 & 672 (limited 168)\\ \hline 
CPU (Intel Core i7 920) & 2.66 & 4 & 42.56\\ 
\noalign{\hrule height 1.5pt}
\end{tabular}
\end{center}
\label{cpu_gpu}
\end{table}
%table%table%table%table%table

\vspace{3mm} Table \ref{node_spec} summarizes the specification of a
computational node used for the experiments.  On each node an Intel Core i7
920 processor \cite{INT09} and a GPU is mounted on a mother board.
Hyper-Threading \cite{HTT} in the Core i7 processor is turned off, and it is
used purely as a four-core CPU.  Compute Capability specifies the version of
hardware level controlled by CUDA, above ver.1.3, which supports double
precision operations.  We used the Intel compiler version 12.0.0 for Fortran/C
codes using options, `-O3' (optimizations including those for loop structures
and memory accesses), `-no-prec-div' and `-no-prec-sqrt' (acceleration of
division and square root operations with slightly less precision),
`-funroll-loops' (unrolling of loops), `-no-fp-port' (no rounding for float
operations), `-ip' (interprocedural optimizations across files), and
`-complex-limited-range' (acceleration for complex variables).  For CUDA we
used the nvcc compiler with options `-O3' and `-arch=sm\_13' (enabling double
precision operations).
%table%table%table%table%table%table%table%table%table
\begin{table}[htdp]
\caption{Setup of a computational node.}
\begin{center}
\begin{tabular}{c|c}
\noalign{\hrule height 1.5pt}
CPU& Intel Core i7 920 2.66 GHz \\ \hline
GPU & NVIDIA GeForce GTX480 $\times$ 1\\ \hline
Motherboard & MSI X58M \\ \hline
Memory & DDR3-10600 2GB $\times$ 6 \\ \hline
OS & Linux Fedora 13 \\ \hline
Fortran/C Compiler &  Intel Fortran/C Composer XE 12.0.0 \\ \hline
CUDA & CUDA version 4.0 \\ 
\noalign{\hrule height 1.5pt}
\end{tabular}
\end{center}
\label{node_spec}
\end{table}
%table%table%table%table%table%table%table%table%table

%%%%%%%%%%%%%%%%%%
\subsection{Memory architecture}
%%%%%%%%%%%%%%%%%%
It is essential in GPGPU coding to design efficiently the parallelized tasks
(termed `threads') to be grouped into subsets with several different classes.
With the variety of memory devices provided in a GPU, see Table
\ref{MemoryList}, each subset has a different `distance' from these devices.
The performance of the GPGPU is critically affected by the choice of theses
subsets because a good design can effectively reduce the memory latency.  In
the present study all of the threads are grouped into `blocks'.  Threads
within a block can share memory devices with fewer latencies.

\begin{figure}[h]
\begin{center}
\includegraphics[width=90mm,angle=0]{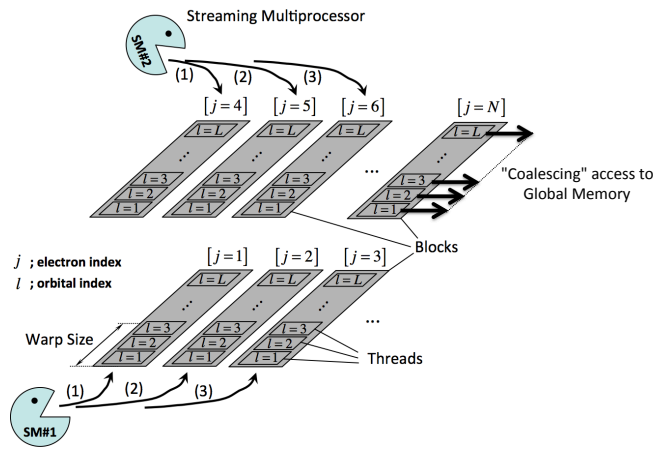}
\end{center}
\caption{A schematic picture showing the relation between blocks, threads, and
streaming multiprocessors (SM).}
\label{packman}
\end{figure}

\vspace{3mm} Each block is assigned to a SM by which the threads within the
block are processed.  The SM processes 32 threads at once, as in vector
processing.  A bunch of 32 threads is called a `warp'.  When a warp accesses
with sequential memory addresses, the latency is much reduced (called
`coalescing').  To conceal the memory latency, the scheduler and dispatcher
for warps monitor which warps are immediately executable (namely which ones
have already completed their memory loads).  Then the scheduled warps are
processed sequentially by the SM.  A schematic picture is shown in Fig.\
\ref{packman}.
%Fig%Fig%Fig%Fig%Fig%Fig%Fig%Fig%Fig%Fig
%Fig%Fig%Fig%Fig%Fig%Fig%Fig%Fig%Fig%Fig

\vspace{3mm} Table \ref{MemoryList} contains only those kinds of
memory relevant to this study, and excludes the texture memory
\cite{CUDA2}.  Off-chip memories are located within the GPU board but
not on the device chip.  They have larger capacities and are
accessible directly from CPU hosts but are in general slower.  On-chip
memories are complementary, namely with higher speed and lower
capacity.  Data required for GPU processing is transferred from the
mother board to off-chip global memory, and is then loaded to on-chip
shared memory, as usual.

\vspace{3mm} The capacity of the global memory ranges from 1 GB to 3 GB,
depending on the product.  As a trade-off against the large capacity it is
about 100 times slower than on-chip memories.  In GTX480, 768 KB of off-chip
L2 cache is available to cover the low speed of the global memory.  Another
off-chip memory with high speed accessibility is the 64 KB constant memory.
Via the constant caches located on every SM the constant memory can be
accessed from all the threads with higher speed, although it is limited to
read-only.  As its name suggests, it is used to store constants referred to by
threads.  On-chip memories inside each SM include registers, shared memories,
and L1 caches.  In GTX480 there are 32,768 registers available for each SM.
Registers are usually used to store the loop index variables defined within
kernel codes, as in the present study.  The 64 KB memory device on each SM can
be shared by all the threads within a block with high speed access.  The 64 KB
capacity is divided into 48 KB and 16 KB parts which work as a shared memory
and L1 cache, respectively.  The user can specify which 48 or 16 KB region
corresponds to the shared memory or L1 cache when the kernel code is compiled.
The access to the global memory refers first to L1 cache and then L2 and
finally to the off-chip global memory device when it fails to load from cache,
which are called cache misses.

\vspace{3mm} Data loading from the global memory takes at least 200 cycles,
and more usually 400 $\sim$ 600 cycles.  To conceal the latency, the GPU
administrates all of the warps and monitors whether it is ready to be executed
with the completion of data load.  With sufficient warps one can ensure that
the processors are almost always executing operations without waiting for data
loads.  To achieve better concealment it is essential to design the code so
that it maintains a large number of warps.  Since it depends on the specs of
each architecture, such as the number of threads per warp, and the maximum
possible number of threads per block {\it etc.}, programming for better
performance requires tuning for each GPU product.
%table%table%table%table%table%table%table%table%table
\begin{table}[ht]
\begin{tabular}{c|c|c|c|c|c}
\noalign{\hrule height 1pt}
& Location & Cache & R/W & Availability & Data maintained \\
\noalign{\hrule height 1pt}
Register & On-chip & - & R/W & within a thread & during a thread \\
\hline
Local memory & Off-chip & L1/L2 &R/W & within a thread & during a thread \\
\hline
Shared memory & On-chip & - & R/W & from all threads & during a block \\
&&&&within a block & \\
\hline
Global memory & Off-chip & L1/L2 & R/W	& from all hosts & during 
host code\\
&&&& and threads & maintains\\
\hline
Constant memory & Off-chip & Yes & R & from all hosts & during 
host code\\
&&&& and threads & maintains\\
\hline
\end{tabular}
\caption{Various kinds of memory in a GPU relevant to this study.
R and W stand for readable and writable, respectively.}
\label{MemoryList}
\end{table}
%table%table%table%table%table%table%table%table%table

%%%%%%%%%%%%%%%%%%%
\section{Coding}
\label{Coding}
%%%%%%%%%%%%%%%%%%%
Only the bottleneck routine for evaluating Eq.\ (\ref{11.12.24.1}) is replaced
by the CUDA kernel code executed on the GPU.  The interface between the main
code in F90 and the CUDA kernel is the same as that in our previous study
\cite{UEJ11}.

%%%%%%%%%%%%%%%%%%
\subsection{Quasi-simultaneous updating}
%%%%%%%%%%%%%%%%
To construct appropriate parallelized degrees of freedom, we
introduced a new scheme for the MC updating of configurations.  Let us
denote a configuration at MC step $\alpha$ by $\vec R^{(\alpha)} =
(\vec r_1^{(\alpha)},\cdots,\vec r_j^{(\alpha)}, \cdots,\vec
r_N^{(\alpha)})$, and consider the update of a particle position $\vec
r_j^{(\alpha)}\rightarrow\vec r^{(\alpha+1)}_j$.  In
configuration-by-configuration updating (simultaneous updating), the
accept/reject of the updating is evaluated based on the ratio of the
probabilities in Eq.\ (\ref{eq05mar12_1}),
%eq%eq%eq%eq%eq%eq%eq
\begin{equation}
\xi _{\rm sim}  
= \frac
{{p\left( {\vec r^{(\alpha+1)}_1 , \cdots ,\vec r^{(\alpha+1)}_{j - 1} ,
\vec r^{(\alpha+1)}_j ,\vec r^{(\alpha+1)}_{j + 1} , 
\cdots ,\vec r^{(\alpha+1)}_N } \right)}}
{{p\left( {\vec r_1^{(\alpha)} , 
\cdots ,\vec r_j^{(\alpha)} , 
\cdots ,\vec r_N^{(\alpha)} } \right)}} \ ,
\label{11.12.23.0a}
\end{equation}
%eq%eq%eq%eq%eq%eq%eq
while in particle-by-particle (PbP, sequential updating),
%eq%eq%eq%eq%eq%eq%eq
\begin{equation}
\xi _{\rm seq}^{\left( j \right)}  
= \frac
{{p\left( {\vec r^{(\alpha+1)}_1 , \cdots ,\vec r^{(\alpha+1)}_{j - 1} ,
\vec r^{(\alpha+1)}_j ,\vec r_{j + 1}^{(\alpha)} , 
\cdots ,\vec r_N^{(\alpha)} } \right)}}
{{p\left( {\vec r^{(\alpha+1)}_1 , \cdots ,\vec r^{(\alpha+1)}_{j - 1} ,
\vec r^{(\alpha)}_j ,\vec r_{j + 1}^{(\alpha)} , 
\cdots ,\vec r_N^{(\alpha)} } \right)}} \ .
\label{11.12.23.0b}
\end{equation}
%eq%eq%eq%eq%eq%eq%eq
The index $j$ in $\xi_{\rm seq}^{\left( j \right)}$ means that the
accept/reject step is made for each particle move, unlike in
configuration-by-configuration updating.  These two updating schemes
give slightly different values for the statistical estimates because
of the different paths of the evaluations, but they coincide with each
other within the statistical errors.  We introduce another updating
scheme based on the ratio
%eq%eq%eq%eq%eq%eq%eq
\begin{equation}
\xi _{\rm q.sim}^{\left( j \right)}  
= \frac
{{p\left( {\vec r^{(\alpha)}_1 , \cdots ,\vec r^{(\alpha)}_{j - 1} ,
\vec r^{(\alpha+1)}_j ,\vec r_{j + 1}^{(\alpha)} , 
\cdots ,\vec r_N^{(\alpha)} } \right)}}
{{p\left( {\vec r_1^{(\alpha)} , 
\cdots ,\vec r_j^{(\alpha)} , 
\cdots ,\vec r_N^{(\alpha)} } \right)}} \ ,
\label{11.12.23.1}
\end{equation}
%eq%eq%eq%eq%eq%eq%eq
termed `quasi-simultaneous updating' (Q.S.).  In this scheme the
accept/reject evaluation for the $j$th particle at step $(\alpha + 1)$
is based on the previously fixed $\alpha$ step configuration and on
each particle position $j$, which gives $N$ individual parallel tasks.

\begin{figure}[h]
\begin{center}
\includegraphics[width=90mm,angle=0]{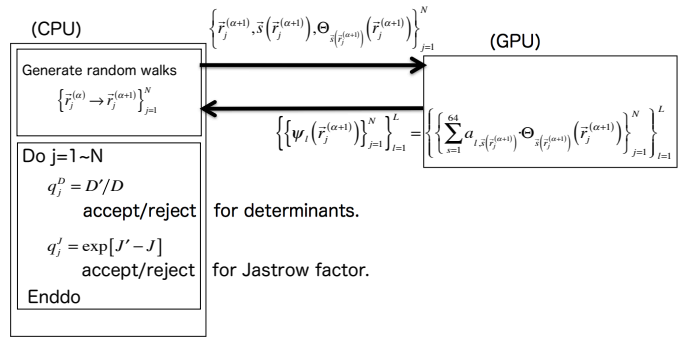}
\end{center}
\caption{Data transfer between a CPU and GPU in the present implementation.
  Trial moves for particle positions are generated and sent to a GPU.  The GPU
  computes updated values of the orbitals to send back to the CPU.  The
  accept/reject evaluation based on the orbital values are performed on the
  CPU.  }
\label{dataTransfer1}
\end{figure}

\vspace{3mm} Evaluating Eq.\ (\ref{11.12.23.1}) reduces to computing the
orbital functions with updated positions, $\left\{\left\{\psi_l \left( {\vec
r^{(\alpha+1)}_j} \right)\right\}_{l=1}^{L}\right\}_{j=1}^{N}$, which requires
$(N\times L)$ independent evaluations.  New trial moves at the $(\alpha + 1)$
step
\begin{equation}
(\vec r_1^{(\alpha)},\cdots,\vec r_j^{(\alpha)},
\cdots,\vec r_N^{(\alpha)})\rightarrow
(\vec r_1^{(\alpha+1)},\cdots,\vec r_j^{(\alpha+1)},
\cdots,\vec r_N^{(\alpha+1)}) \ ,
\label{11.12.23.2}
\end{equation}
are generated on a CPU and sent to a GPU and the orbital functions are
evaluated (see Fig.\ \ref{dataTransfer1}).  For TiO$_2$
(4$\times$4$\times$4) this gives $N\times L =1536 \times 384 =
589,824$ parallelized tasks.  Note that the parallelized multiplicity
$(N\times L)$ scales as $N^2$ since the number of orbitals $L \propto
N$.  The concealment of memory latency is more efficient when the
multiplicity increases, and hence we expect better performance for
larger systems.
%fig%fig%fig%fig%fig%fig
%fig%fig%fig%fig%fig%fig

%%%%%%%%%%%%%%%%%%
\subsection{Assigning blocks}
%%%%%%%%%%%%%%%%
Each thread evaluating $\psi_l \left( {\vec r^{(\alpha+1)}_j} \right)$
is labelled by $(j,l)$.  As mentioned in \S II.B, the memory access
for the coalescing should take $l$ as the sequential index for the
data load of $a_{l,s\left( {\vec r_j } \right)} = a\left[ {l,j_x ,j_y
    ,j_z } \right]$, as required for Eq.\ (\ref{11.12.24.1}).  We
therefore assigned the threads sharing the same $j$ with sequentially
varying $l=1\sim L$ within a block for the coalescing.  The number of
orbitals, $L$, cannot therefore exceed the maximum possible number of
threads within a block, which is 1,024 for GTX480.  The number $L$ can
usually be reduced to $L<N$ using the symmetry of the system.  
It is then a natural choice 
{\color{black}
to assign $L$ within a block
rather than $N$, which is always being larger than $L$.  
}
For non-magnetic solid systems as
in the present work, $L$ is reduced to $\sim N/4$ by the symmetries,
giving the limit $N \lesssim$ 4,096 for GTX480.  For magnetic systems,
$L \sim N/2$, and hence $N \lesssim$ 2,048, and for systems without
time-reversal symmetry it becomes 1,024.  This limitation is
consistent with the maximum simulation size of contemporary QMC
electronic structure calculations, which are able to treat at most a
few thousand electrons in extended systems \cite{MAE10}.  Furthermore
the limit of 1,024 in GTX480 is expected to double in future
architectures \footnote{In GTX275 used in our previous work
  \cite{UEJ11} it was 512 for the limitation of threads within a
  block.}, making this issue less important.

\vspace{3mm} When evaluating Eq.\ (\ref{11.12.24.1}), all the threads within a
block refer to the same $\Theta _s \left( {\vec r_j }\right)$ with fixed $j$.
Once a warp loads $\left\{\Theta _s \left( {\vec r_j } \right)
\right\}_{s=1}^{64}$, these data are stored in L1 and L2 cache including their
neighboring data.  We can then expect effective cache hits for the data load.
The operation of each thread, 64 terms multiply and add, easily fits the FMA
of the GPU.

%%%%%%%%%%%%%%%%%%%%%%%%
\subsection{Other code prepared for comparisons}
%%%%%%%%%%%%%%%%%%%%%%%%
We prepared several other versions of the code with different updating
schemes and thread/block assignments, in order to compare the
performance.  The original CPU implementation provided by the CASINO
distribution \cite{NEE10} is the particle-by-particle (PbP) algorithm,
with double precision updating, which we refer to as [(0a) CPU/PbP].
Another version, [(1) GPU/PbP], is the GPU version of (0a) but with
single precision updating, which is useful for studying deviations
between single and double precision computations.  In this version the
GPU kernel is called with a single fixed $j$ (PbP).  Each term,
$a_{ls}\cdot \Theta(\vec r_j)$ in the summation of Eq.\
(\ref{11.12.24.1}), labelled by $s$, is calculated by each thread, the
sum of which is obtained by the reduction operation \cite{CUDA2} among
all threads.  The threads indexed by $s$ are grouped into those
sharing the same $l$ and hence the blocks are labelled by the orbital
index $l$.  Another reference is the version [(2)
GPU/Q.S./non-coalescing], which uses quasi-simultaneous (Q.S.)
updating, but the same thread/block assignment as the version (1),
namely each thread calculates only the product $a_{ls}\cdot
\Theta(\vec r_j)$.  In this case the blocks are labelled by $(j,l)$.
Coalescing does not work in this version.  The indices for threads and
blocks are summarized in Table \ref{kernel_time}.  
Comparing (1) and (2) shows firstly 
{\color{black} 
how the performance in speed is improved by the simultaneous
data transfers for $j = 1\sim N$ in Q.S. compared to
the sequential transfers in PbP.
}  
Secondly we can see
how much the energy deviates due to using Q.S.\ updating instead of
PbP.  Our final implementation, described in \S IV.B, is termed [(3)
GPU/Q.S./coalescing].  The comparison between (1) and (2) is within
the single precision treatment.  To compare the Q.S.\ scheme in single
and double precision we also prepared [(0b) CPU/Q.S.], which is a
double precision version of Q.S.

%%%%%%%%%%%%%%%%%%%%%%%%
\section{Results}
%%%%%%%%%%%%%%%%%%%%%%%%

%%%%%%%%%%%%%%%%%%%%%%%%
\subsection{Acceleration performance}
%%%%%%%%%%%%%%%%%%%%%%%%
Tables \ref{kernel_time} and \ref{kernel_time0} summarize the acceleration
factors and computational time taken for the bottleneck kernel part within a
MC step.  The results are evaluated for systems with $N$ = 216 (Si, $3\times
3\times 3$) and 1,536 (TiO$_2$, $4\times 4\times 4$).  A better performance
was obtained using implementation (3) rather than (2) and (1), and with larger
system sizes $N$.
%table%table%table%table%table%table%table%table%table%table
\begin{table}[ht]
\caption{Comparison of acceleration factors for each implementation evaluated
    from Table \ref{kernel_time0}.  `PbP' and `Q.S.' stand for the
    particle-by-particle and quasi-simultaneous updating schemes,
    respectively.  The `Index' column shows which indices in Eq.\
    (\ref{11.12.24.1}) are assigned to threads and blocks in the GPU.  For
    system sizes $N$, see Table \ref{system_size} for a more detailed
    description of the systems.  }
\begin{center}
\begin{tabular}{l|c|c|c|c}
\noalign{\hrule height 1.5pt}
 &  \multicolumn{2}{c|}{Index} & \multicolumn{2}{c}{acceleration factor} \\
 &  block & thread & $N=$ 216 & $N=$ 1,536  \\
\noalign{\hrule height 1.5pt}
(1) GPU/PbP & $l$ & $s$ & 0.41 & 1.47 \\ \hline
(2) GPU/Q.S./non-coalescing & $l,j$ & $s$ & 6.39 & 5.61 \\ \hline
(3) GPU/Q.S./coalescing & $j$ & $l$ &16.58 & 30.67 \\ 
\noalign{\hrule height 1.5pt}
\end{tabular}
\end{center}
\label{kernel_time}
\end{table}
%table%table%table%table%table%table%table%table%table%table
The improvement from (1) to (2) is attributed to the increased number
of threads in Q.S.\ due to processing the particle indices $j=1\sim N$
simultaneously.  This also brings about improved efficiency in the
data transfer between the CPU and GPU, which is simultaneous transfer
in version (2) and repeated transfers for each $j$ in PbP version (1).
The improvement from (2) to (3) arises because the number of
operations performed on each thread is increased; multiply and add
summation with 64 terms in (3) and just one multiplication in (2).
The fact that memory coalescing only works in (3) also contributes to
the improvement, for which a detailed analysis will be given in \S
VI.C.
%table%table%table%table%table%table%table%table%table%table
\begin{table}[ht]
  \caption{Comparison of the computational times (ms) per Monte Carlo step in
    each implementation.  `PbP' and `Q.S.' stand for the particle-by-particle
    and quasi-simultaneous updating schemes, respectively.  More information
    about the systems (of sizes $N$) are given in Table \ref{system_size}.  }
\begin{center}
\begin{tabular}{l|c|c|c|c}
\noalign{\hrule height 1.5pt}
 &  \multicolumn{2}{c|}{$N=$ 216} & \multicolumn{2}{c}{$N=$ 1,536} \\
 &  CPU & GPU & CPU & GPU \\ 
\noalign{\hrule height 1.5pt}
(1) GPU/PbP & & 6.76 & & 68.03 \\ \cline{1-1}\cline{3-3}\cline{5-5}
(2) GPU/Q.S./non-coalescing & 2.77 & 0.43 & 100.00 & 17.83 \\  \cline{1-1}\cline{3-3}\cline{5-5}
(3) GPU/Q.S./coalescing & & 0.17 & & 3.26 \\
\noalign{\hrule height 1.5pt}
\end{tabular}
\end{center}
\label{kernel_time0}
\end{table}
%table%table%table%table%table%table%table%table%table%table

%%%%%%%%%%%%%%%%%%%%%%%%
\subsection{Deviation in energy values}
%%%%%%%%%%%%%%%%%%%%%%%%
In GPGPU for scientific simulations it is important to consider
whether the deviation in the results due to the single precision
operations are within the required accuracy.  For the present
electronic structure simulation the deviation should be within $\Delta
E \sim$ 0.001 [hartree] in the energy estimation, known as the {\it
  chemical accuracy}.  Table \ref{energy_eval} shows the results from
each implementation, the ground state energy of Si, $3\times 3\times
3$ ($N$ = 216), by 1,000,000 MC steps with sampling every 10 MC steps
(`dcorr' = 10, see \S II.C).
%table%table%table%table%table%table%table%table%table%table
\begin{table}[ht]
  \caption{ The ground state energy of Si, $3\times 3\times 3$ ($N$ = 216), from
    1,000,000 MC steps, sampling at every 10 steps, evaluated by each
    implementation.  `PbP' and `Q.S.' stand for the particle-by-particle and
    quasi-simultaneous updating schemes, respectively.  }
\begin{center}
\begin{tabular}{l|c}
\noalign{\hrule height 1.5pt}
&Energy (hartree/primitiveCell)  \\ 
\noalign{\hrule height 1.5pt}
(0a) CPU/PbP (double prec.) & -7.9590865 $\pm$ 0.0001749 \\ \hline
(1) GPU/PbP (single prec.) & -7.9589381 $\pm$ 0.0001754 \\ \hline
(0b) CPU/Q.S.  (double prec.)& -7.9591560 $\pm$ 0.0001755 \\ \hline
(2,3) GPU/Q.S. (single prec.) & -7.9592106 $\pm$ 0.0001698 \\ 
\noalign{\hrule height 1.5pt}
\end{tabular}
\end{center}
\label{energy_eval}
\end{table}
%table%table%table%table%table%table%table%table%table%table
The comparison between (0a) and (1) gives the deviation due to the
change from double to single precision evaluation.  The deviation is
within the statistical error bars, but the agreement is poorer than
single precision, which is expected to be correct to around six
digits.  To understand this we must remember that the energies given
in Table \ref{energy_eval} are statistical estimates obtained from
different accept/reject paths after 1,000,000 MC steps.  When we look
at the difference after a single MC step, agreement is confirmed
within six digits, as shown in Table \ref{energy_eval_1step}.  Even
such small deviations may give rise to different decisions along a
accept/reject branch.  Once a different decision occurs, the
subsequent random walk takes different paths, giving different
estimates which are outside of the six digits but within the
statistical error bars.
%table%table%table%table%table%table%table%table%table%table
\begin{table}[ht]
  \caption{ Comparison between the energies after a MC step
    evaluated in each implementation.  `PbP' stands for the 
    particle-by-particle updating scheme.  }
\begin{center}
\begin{tabular}{c|c}
\noalign{\hrule height 1.5pt}
&Energy (hartree)  \\ 
\noalign{\hrule height 1.5pt}
(0a) CPU/PbP (double prec.) &  -7.95075065699 \\ \hline
(1) GPU/PbP (single prec.)&  -7.95075065360 \\ 
\noalign{\hrule height 1.5pt}
\end{tabular}
\end{center}
\label{energy_eval_1step}
\end{table}
%table%table%table%table%table%table%table%table%table%table

\vspace{3mm} A comparison of (0a) and (0b) in Table \ref{energy_eval}
gives the deviation due purely to the two different updating schemes.
As expected it is confirmed that these energies agree to within the
statistical error bars.  The result of (2,3) includes both the
deviation due to the changes in the updating scheme and the numerical
precision.  Comparing with the reference (0a) demonstrates that our
updating scheme keeps the result within the required accuracy.

\vspace{3mm} %We note that we can insist the confirmation of the
%deviation being within statistical error bars only to some limited extent.  
In the present study, the updated orbital values from the GPGPU/single
precision evaluations are used only to determine if updated particle
positions are accepted or rejected.  The energies reported in Table
\ref{energy_eval} were calculated using CPU/double precision.  The
difference between single and double precisions alters the paths of
the random walks and hence where the $\vec R$ space is sampled.  If
the energies themselves were also evaluated using single precision it
might introduce significant biases, but we have not investigated this
here.  As mentioned in \S VI.d, there are further possibilities for
accelerating the second largest bottleneck, which is the updating of
the Slater determinants (or equivalents) using GPGPU.  In this scheme
the updated single precision value of the many-body wave function is
used to evaluate the energies, and the errors in these energies should
be considered carefully to make sure the results are still within
chemical accuracy.

%%%%%%%%%%%%%%%%%%%%%%%%
\subsection{System size dependence of the performance}
\begin{figure}[htdp]
\begin{center}
\includegraphics[width=100mm]{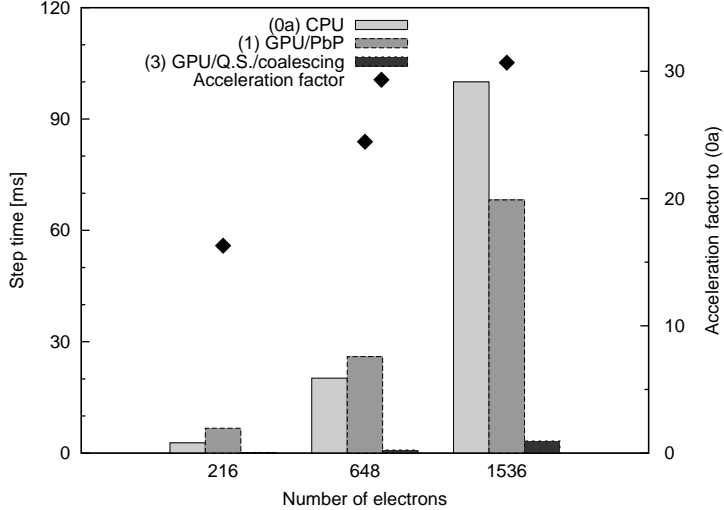}
\caption{Computational time [ms] per MC step and corresponding
  acceleration factors for system sizes $N$ (number of electrons).
  `PbP' and `Q.S.' stand for the particle-by-particle and
  quasi-simultaneous updating schemes, respectively.  }
\label{time}
\end{center}
\end{figure}

%%%%%%%%%%%%%%%%%%%%%%%%
The acceleration factors achieved by implementation (3) applied to
$N=$ 216 (Si/$2 \times 2 \times 2$), 648 (TiO$_2$/$3 \times 3 \times
3$), and 1,536 (TiO$_2$/$4 \times 4 \times 4$) are summarized in Table
\ref{system_size}.  Figure \ref{time} also shows the acceleration
factors and computational times (ms) taken for a MC step in (0a), (1),
and (3).
%fig%fig%fig%fig%fig%fig%fig%fig%fig%fig%fig%fig%fig%fig%fig
%fig%fig%fig%fig%fig%fig%fig%fig%fig%fig%fig%fig%fig%fig%fig
As mentioned in \S IV.A, the number of parallelized threads scales as
$(N \times L) \sim N^2$, so that the efficiency of the GPU improves
for larger systems.  Coalescing in implementation (3) also becomes
more effective for larger $L$.  However there is a drawback for larger
systems in the data transfer costs, especially for returning $\left\{
  {\left\{ {{\psi _l}\left( {{{\vec r}_i}} \right)} \right\}_{i =
      1}^N} \right\}_{l = 1}^L$ from the GPU to CPU.  
As overall cancellation, the acceleration seems to scale as $\sim
N$
{\color{black}, rather than $ \sim N^2$ (the number of threads). 
}
%table%table%table%table%table%table%table%table%table%table
\begin{table}[ht]
  \caption{Data transfer costs between the CPU and GPU
    from implementation (3) in Table \ref{kernel_time}.
    The dependence on  the system size $N$ (number of electrons) is shown.
  }
\begin{center}
\begin{tabular}{c||c|c|c|c|c}
\noalign{\hrule height 1.5pt}
 & \multicolumn{2}{c|}{CPU $\to$ GPU ($\mu$s)} & GPU $\to$ CPU  ($\mu$s)
& Total  & Percentage \\
$N$ 
& $s\left(\vec r'_j \right)$ & $\Theta_{s\left(\vec r'_j \right)} \left(\vec r'_j \right)$ 
& $\psi_l \left(\vec r'_j \right)$& ($\mu$s)& within total GPU time
\\ \noalign{\hrule height 1.5pt}
216   & 16.8   & 19.5 & 29.1 & 71.9 & 47.6 \%\\ \hline
648   & 18.8 & 32.0 & 290.9 &405.1& 45.8 \%\\ \hline
1,536  & 25.5 & 131.1 & 1046.6 & 1203.2 & 36.9 \%\\ 
\noalign{\hrule height 1.5pt}
\end{tabular}
\end{center}
\label{cuda_memcopy}
\end{table}
%table%table%table%table%table%table%table%table%table%table
Table \ref{cuda_memcopy} summarizes the data transfer times using
ver.\ (3), which show that the transfer cost increases with system
size.  The percentage of the total GPU time is, however, decreased
because the number of GPU operations also increases.

\vspace{3mm} Another remarkable fact illustrated by Fig.\ \ref{time}
is that the performance of ver.\ (1) is inferior up to $N$ = 648 but
superior for larger $N$ = 1,536.  The only possible reason for that is
the concealment of memory latency, because in this implementation
there is no coalescing and the operation load on each thread is simply
a multiplication, although it has the largest number of warps among
all the implementations.

%%%%%%%%%%%%%%%%%%%%%%%%
\section{Discussions}
%%%%%%%%%%%%%%%%%%%%%%%%
To prevent redundancy we omit discussions of the evaluation of how
much of the original CPU code is optimized, and of how we interpret
the acceleration factors achieved in terms of the actual usefulness of
GPGPU, as these were discussed in our previous report \cite{UEJ11}.

%%%%%%%%%%%%%%%%%
\subsection{Acceleration performance}
%%%%%%%%%%%%%%%%%
The ideal limit of the acceleration factor to be compared with our
achievement of 30.7 is that of (1345/10.64)=126.4, where 1345 [GFlops]
is the ideal limit for the GPU discussed in \S III.A and 10.64 is the
single core performance of the Intel Core i7 processor used here for
implementation (0a).  This limit might be achieved when the ratio of
the cost of memory loads to that for operations approaches zero,
although this is not possible in practice.  
{\color{black}
This ratio corresponds to those shown in the last column of Table
\ref{cuda_memcopy} as in percentages.}
As shown in the table,
the ratio is decreased for larger systems and
hence we expect better performance. 

\vspace{3mm} As another evaluation of our achievement, we estimated
the FLOPS of our implementations applied to TiO$_{2}$ ($N$=1,536) as
listed in Table \ref{perfomance}.  The values are obtained from the
number of operations required only for Eq.\ (\ref{11.12.24.1}) [(2
operations) $\times$ (2 components in complex numbers) per term]
divided by the time taken for the GPU kernel execution.  We did not
take into account the operations required to identify which subset
$\left\{s\right\}_{s=1}^{4^3} \subset \left\{\tilde s\right\}_{\tilde
  s = 1}^{S=50^3\sim 60^3}$ should be chosen to form the coefficients
$\left\{a_{ls}\right\}$.  The actual FLOPS should therefore be larger
than those given in the table.
%table%table%table%table%table%table%table%table%table%table
\begin{table}[h]
  \caption{Estimated FLOPS for each implementation and the ratios to the ideal
    performances.  `PbP' and `Q.S.' stand for the particle-by-particle and
    quasi-simultaneous updating schemes, respectively.  }
\begin{center}
\begin{tabular}{l|r|r}
\noalign{\hrule height 1.5pt}
 & Performance (GFlops) & Ratio to Ideal performance \\ 
\noalign{\hrule height 1.5pt}
(0a) CPU/PbP & 1.50 & 14.10 \% \\ \hline
(1) GPU/PbP & 4.62 & 0.34 \% \\ \hline
(2) GPU/Q.S./non-coalescing  & 9.05 & 0.71 \% \\ \hline
(3) GPU/Q.S./coalescing & 73.98 & 5.50 \% \\ 
\noalign{\hrule height 1.5pt}
\end{tabular}
\end{center}
\label{perfomance}
\end{table}
%table%table%table%table%table%table%table%table%table%table
In this evaluation, our achievement of a factor 30.7 gives an
efficiency of only 5.5\% of the ideal performance.  For reference, the
CUBLAS (GPGPU of BLAS Level 3) is known to give 400 GFlops on the
NVIDIA Tesla C2050, which is 38.8\% of the peak performance
\cite{tesla-blas}.  The reason for the lower percentage in our case is
the smaller number of operations per thread, 64 terms multiply and add
summations for (3), and just a single multiplication for (1) and (2).
The amount does not depend on the system size because of the use of a
localized spline basis set, although this is a disadvantage for GPGPU
in the sense that the number of operations is smaller.

%%%%%%%%%%%%%%%%
\subsection{Performances in memory access}
%%%%%%%%%%%%%%%%
Table \ref{prof_opt} summarizes the memory load performances of each
implementation as measured by `Compute Visual Profiler' for CUDA
\cite{profiler}.
%table%table%table%table%table%table%table%table%table%table
\begin{table}[ht]
  \caption{Performance in global memory access for each implementation.
    `PbP' and `Q.S.' stand for the particle-by-particle and quasi-simultaneous
    updating schemes, respectively.  }
\begin{center}
\begin{tabular}{l|r|r|r}
\noalign{\hrule height 1.5pt}
 & \multicolumn{2}{c|}{Global Memory Access} & SM activity \\
 &Num. of 32bit Load& Throughput (GB/s)&   \\ 
\noalign{\hrule height 1.5pt}
(1) GPU/PbP & 24,576 & 13.5 & 88.1 \%  \\ \hline
(2) GPU/Q.S./non-coalescing & 188,744,000 & 18.1 & 99.4 \% \\ \hline
(3) GPU/Q.S./coalescing & 7,077,890& 153.0 &100.0 \% \\ 
\noalign{\hrule height 1.5pt}
\end{tabular}
\end{center}
\label{prof_opt}
\end{table}
%table%table%table%table%table%table%table%table%table%table
The increase in the number of 32 bit load from (1) to (2) is simply
because of the increased number of threads due to the simultaneous
processing with respect to the number of particles $N$ in Q.S.  From
the coalescing in (3) we see a remarkable reduction in the number of
memory loads by a factor of $\sim$ 27.  Correspondingly the throughput
becomes much closer to the peak value, 177.4 GB/s, compared with the
poor achievements in (1) and (2) due to the random access of
$\left\{a_{js}\right\}$.

\vspace{3mm} The SM activity in Table \ref{prof_opt} is a measure of
the efficiency of the concealment of the memory latency.  This
quantity is defined as the ratio of the number of cycles at which the
operations started after the completion of memory loads to the total
number of cycles taken for the kernel execution.  With efficient
concealing, at least one of the threads is always ready for execution
at each cycle, and hence this quantity is expected to be close to
100\%.  Since the concealment becomes more effective with larger
numbers of warps, the SM activity is increased from (1) to (2) by the
increased number of threads.  Though the number is decreased again in
(3) by a factor of $\sim$ 64, greatly accelerated memory accesses by
the coalescing improve the SM activity to 100\%.

%%%%%%%%%%%%%%%%
\subsection{Shared memory and read only memory}
%%%%%%%%%%%%%%%%
In our previous work \cite{UEJ11} we found that it was quite effective
to exploit high-speed read-only memory.  In the present work we have
investigated further improvements by exploiting high-speed read-only
memory but found no significant gains.  This is summarized as follows:
(i) in the present case the data required for the operation on each
thread is large and does not fit into the high-speed memory devices,
and (ii) even without explicit use of high-speed devices, the compiler
automatically assigns them to L1 cache, and the explicit data transfer
to shared memory by the user gives rather slower performance than
automatic assignment.

\vspace{3mm} The data loads considered in the present case are
$\left\{ {{{\tilde a}_{l,\tilde s}}} \right\}_{\tilde s = 1}^{{S \sim
    250,000}}$, which is initially stored in the global memory.  After
choosing the subset $\left\{\left\{ {a_{l,s} } \right\}_{s =
    1}^{64}\right\}_{l=1}^{L}$ from $\left\{ {{{\tilde a}_{l,\tilde
        s}}} \right\}$, our best implementation (3) loads them by
coalescing access to the global memory.  However, the access speed to
the on-chip shared memory is around 100 times faster than
that. \footnote{With coalescing the maximum speed for the global
  memory access is expected being around 200 clock cycles, compared
  with 2 clock cycles for on-chip memories.}.  A block sharing a
shared memory device has a common $j$ so the set $\left\{\left\{
    {a_{l,s} } \right\}_{s = 1}^{64}\right\}_{l=1}^{L}$ is shared by
all the threads within the block.  The 64 KB capacity of the device
corresponds to 16,000 (= 64 KB/4B) elements of single precision data.
It is therefore possible to accommodate $\left\{ {a_{l,s} } \right\}$
when $L < 16,000/64 = 250$, so that the total number of orbitals is
less than 250.  Table \ref{system_size} shows that Si ($3\times
3\times 3$) and TiO$_2$ ($3\times 3\times 3$) correspond to this case.
Storing the data within the on-chip memories becomes advantageous when
the data is referenced repeatedly by the warps.  The number of
repeated references is given in our case by the ratio of $L \sim$ 250
to the warp size, 32, which is less than 10.  Beyond this number of
repeats the SM switches to another process with different $j$ and
hence the corresponding new data for $\left\{ {a_{l,s} } \right\}$
should be loaded again from the global memory.  We have tried such an
implementation using shared memory devices but we did not find any
improvements over (3), possibly because of the small number of
repeated references.  Such an improvement would already be included
implicitly in (3) by L1 cache acting on the shared device.  The
explicit usage of the device as the shared memory seems to reduce the
performance.

\vspace{3mm} Another choice is to use constant memory.  Unlike shared
memory it is located off chip and blocks with different $j$ indices
pick up their subset $\left\{ {a_{l,s} } \right\}$.  The memory should
therefore be able to provide the whole set of $\left\{ {{{\tilde
        a}_{l,\tilde s}}} \right\}$, which is far beyond the size of
the constant memory and is therefore infeasible.  
{\color{black} Another data required for the}
GPU operation is
$\left\{\left\{ {\Theta _s \left( {\vec r_j } \right)} \right\}_{j =
    1}^N \right\}_{s=1}^{64}$ in Eq.\ (\ref{11.12.24.1}), the total
size of which can be accommodated within the constant memory for $N <$
250.  Our trial implementation using constant memory for
$\left\{\Theta_s \right\}$ actually gives a slight improvement in
performance, by $\sim$ 4\%, but this is only applicable to smaller
system sizes.  The data $\left\{ {\Theta _s \left( {\vec r_j^{(k)} }
    \right)} \right\}_{j = 1}^N $ is updated at every MC step and
hence the {\it life time} in cache is short, giving only a slight
improvement in performance.

%%%%%%%%%%%%%%%%
\subsection{Acceleration of Slater Determinant Updating}
%%%%%%%%%%%%%%%%
The bottleneck operation of updating of orbital functions, which are
computed by the GPU, gives a system size dependence of $O(N^2)$
\cite{NEE10}.  The corresponding updating of the many-body wave
function, (\ref{eq5}), actually takes $O(N^2)$ in the PbP
implementation \cite{HAM94,CEP77}, in spite of the fact that
evaluating a determinant scales as $O(N^3)$.  In PbP, say $\vec r_j
\to \vec r'_j$, only one column of the determinant including $\vec
r'_j$ is updated at each step.  The updating of the determinant can
hence be evaluated based on the Laplace expansion with respect to the
column, and the other co-factors are unchanged.  This algorithm, known
as Sherman-Morrison updating \cite{CEP77} therefore requires only
$O(N^2)$ operations \cite{NEE10}.  In our Q.S., the updated orbitals
are stored in an array and are read sequentially to update the
determinant using this algorithm.  The cost of the updating of the
many-body wave function amounts to 19.4\% of the time compared to 39.5
\% for the present bottleneck, the orbital updating, when $N$=1,536,
dcorr=1.  The ratio of the former to the latter increases with larger
dcorr.  This operation mostly involves BLAS routines (ddot and daxpy)
which could be replaced by CUBLAS in future work.

%%%%%%%%%%%%%%%%
\subsection{As a prototype of linear algebra problems}
%%%%%%%%%%%%%%%%
The evaluation of Eq.\ (\ref{11.12.24.1}) can be written in terms of
linear algebra,
%eq%eq%eq%eq%eq%eq%eq
\begin{equation}
\label{12.1.20.3}
\left( {\begin{array}{*{20}{c}}
{\psi _1\left( {{{\vec r}_j}} \right)}\\
{\psi _2\left( {{{\vec r}_j}} \right)}\\
 \vdots \\
{\psi _L\left( {{{\vec r}_j}} \right)}
\end{array}} \right) 
= A\left( {{{\vec r}_j}} \right)\cdot\left( {\begin{array}{*{20}{c}}
{{\Theta _1}\left( {{{\vec r}_j}} \right)}\\
{{\Theta _2}\left( {{{\vec r}_j}} \right)}\\
 \vdots \\
{{\Theta _{64}}\left( {{{\vec r}_j}} \right)}
\end{array}} \right) \ ,
\end{equation}
%eq%eq%eq%eq%eq%eq%eq
for which the matrix $A$ is defined by
%eq%eq%eq%eq%eq%eq%eq
\begin{equation}
\label{12.3.27.1}
A = \left\{ {{a_{ls}}} \right\} 
= \left\{ {{a_{l,s\left( {{{\vec r}_j}} \right)}}} \right\} = A\left( {{{\vec r}_j}} \right) \ .
\end{equation}
%eq%eq%eq%eq%eq%eq%eq
If the matrix $A$ is constant, we can write Eq.\ (\ref{12.1.20.3}) as
a matrix multiplication:
%eq%eq%eq%eq%eq%eq%eq
\begin{equation}
\label{12.1.20.2}
\left( {\begin{array}{*{20}{c}}
{\psi _1^{}\left( {{{\vec r}_1}} \right)}&{\psi _1\left( {{{\vec r}_2}} \right)}& \cdots &{\psi _1\left( {{{\vec r}_N}} \right)}\\
{\psi _2^{}\left( {{{\vec r}_1}} \right)}& \ddots & & \vdots \\
 \vdots &{}&{}&{}\\
{\psi _L^{}\left( {{{\vec r}_1}} \right)}& \cdots &{}&{\psi _L\left( {{{\vec r}_N}} \right)}
\end{array}} \right) = A\cdot\left( {\begin{array}{*{20}{c}}
{{\Theta _1}\left( {{{\vec r}_1}} \right)}&{{\Theta _1}\left( {{{\vec r}_2}} \right)}& \cdots &{{\Theta _1}\left( {{{\vec r}_N}} \right)}\\
{{\Theta _2}\left( {{{\vec r}_1}} \right)}& \ddots &{}& \vdots \\
 \vdots &&&\\
{{\Theta _{64}}\left( {{{\vec r}_1}} \right)}& \cdots &{}&{{\Theta _{64}}\left( {{{\vec r}_N}} \right)}
\end{array}} \right) \ ,
\end{equation}
%eq%eq%eq%eq%eq%eq%eq
and using this we could increase the ratio of operations to data
transfer, and hence the efficiency of GPGPU by a large amount.  Fig.\
\ref{coefficient}, however, reminds us that the matrix $A$ in Eq.\
(\ref{12.3.27.1}) is randomly varying {\color{black} in its elements choice} indexed by $\vec r_j$, but
within the given constant elements $\left\{ {{{\tilde a}_{l,\tilde s}}} \right\}$, 
and hence 
{\color{black}
we cannot reduce Eq. (\ref{12.1.20.3}) into a unified form as Eq.\ (\ref{12.1.20.2}).} 

\vspace{3mm} Since our calculation of Eq.\ (\ref{11.12.24.1}) is
linear algebra we considered whether it could be performed efficiently
using CUBLAS.  To use CUBLAS we have to construct $A\left( {{{\vec
        r}_j}} \right)$ on the CPU at every MC step, transfer it to
the GPU, and then call CUBLAS.  In our case the randomly varying
matrix $A$ is not large, however, the cost of constructing it cannot
be compensated by the high performance of CUBLAS.  Since the enlarged
matrix $\tilde A = \left\{ {{\tilde a_{l \tilde s}}} \right\}_{\tilde
  s = 1}^{{S}}$ is a constant matrix, only one data transfer to the
GPU is required in principle, but this is not possible with CUBLAS
which requires the operands to be transferred every time for the
operations.  In this sense, our achievement here corresponds to an
effective GPGPU implementation of the algorithm for the prototype as
explained above, Eq.\ (\ref{12.1.20.3}) with a randomly chosen matrix.

%%%%%%%%%%%%%%%%%%%%%%%%
\section{Concluding Remarks} 
\label{section:concluding_remarks}
%%%%%%%%%%%%%%%%%%%%%%%%
We have applied GPGPU to the evaluation of the orbital functions in
{\it ab-initio} Quantum Monte Carlo electronic structure calculations,
which we identified as the computational bottleneck.  For efficiency
we propose a new updating scheme for generating trial moves for the
walkers in the Monte Carlo sampling, which we call quasi-simultaneous
updating.  Using this scheme we achieved a speedup of more than a
factor of 30 compared with using a single core CPU.  The GPGPU
implementation gives a deviation in energy from original CPU
evaluation which is smaller that the required chemical accuracy.
Though the effective performance in operations amounts to 74 GFlops,
which is only 5.5\% of the peak performance, the memory throughput
reaches 153 GB/s, which is 86\% of the peak value with almost perfect
concealment of memory latency as shown by the SM activity.  The
implementation presented here is a prototypical problem of linear
algebra with a sort of random matrix, processed by GPGPU.

%SSSSSSSSSSSSSSSSSSSSSSSS
\section{Acknowledgments}
%SSSSSSSSSSSSSSSSSSSSSSSSSSSSSSSSSSSSSSS
RM is grateful for financial support provided by a Grant in Aid for
Scientific Research on Innovative Areas ``Materials Design through
Computics: Complex Correlation and Non-Equilibrium Dynamics'' (Grant
No.\ 22104011), and ``Optical Science of Dynamically Correlated
Electrons'' (Grant No.\ 23104714) of the Ministry of Education,
Culture, Sports, Science, and Technology (KAKENHI-MEXT/Japan).
The authors would like to express our special thanks to Richard J. Needs 
and Kenta Hongo for their useful comments and kind supports for us.

%%% References %%%%%

%\newpage
\end{document}